# Method Study on the 3x+1 Problem

Du Lizhi
(College of Computer Science and technology, Wuhan University of Science and Technology, Wuhan,430065, telephone: 86 13554171855)

**Abstract:** The 3x+1 problem is one of the most classical problems in computer science, related to many fields. As it is thought by scientists a highly hard problem, resolving it successfully not only can improve the research in many relating fields, but also be meaningful to the method study. By deep analyzing the 3x+1 calculation process with the input positive integer becoming greater, we find a useful way for solving this problem with high probability. By making use of the greater calculating ability of great computers and the internet, our way is a valid and powerful way for utterly solving the 3x+1 problem. This way can be expressed in three points: 1) If we can find a positive integer N, for any positive integer less than $2^N$, the times of dividing 2 out of its stopping time is less than or equal to N, then the 3x+1 conjecture is true; 2) This N may be big, so the calculation may be too big. Our way for solving this is: to find a positive integer K, for all positive integers less than $2^K$, not all the times of dividing 2 out of their stopping time for these integers are less than or equal to K, some part of these are greater than K, but the number of this part becomes less and less with the K increasing; 3) This K and the calculation may also be too big, our way for solving this is: to find a positive integer R, for all positive integers less than $2^R$, as above, out of their stopping time, the times of dividing 2 of some part of these integers are greater than R, also the number of this part integers does not become less immediately with the R increasing, but the increasing rate of this number is less and less until to below zero.

Next, I use Chinese to explain my method carefully.

# 解决3x+1问题方案研究

杜立智
(武汉科技大学计算机科学与技术学院　武汉 430065)

**摘要:** 3x+1问题是计算机科学领域最经典的问题之一，牵涉到许多研究领域，同时由于该问题被权威专家判定为超难，它的成功解决不仅会对其相关领域的研究产生推动促进，更可能意味着方法论上的重大突破。本文通过深入分析3x+1问题随着其输入自然数的不断扩大所蕴含的演化规律，给出了一条有力可行的解决该问题的方案。随着大型计算机以及互联网计算功能的应用，本方案可能直接导致决定性结果。我们的方案分三个层次: 1）若能找到某个自然数N，对于所有小于$2^N$的自然数，它们的停止时间偶变换的次数均不超过N，则3x+1猜想成立; 2）这样的N可能较大，导致计算量超大，解决办法是: 找到某个自然数K，对于所有小于$2^K$的自然数，它们的停止时间偶变换的次数有一部分超过了K,但这一部分的个数随着K的增加而逐步衰减;3）这样的K依然可能较大，导致计算量超大，解决办法是: 找到某个自然数R，对于所有小于$2^R$的自然数，它们的停止时间偶变换的次数有一部分超过了R，这一部分的个数并没有随着R的增加而立即衰减，但其增加的趋势在衰减,并能衰减到零以下。

## 1 引言

计算机算法和计算复杂性理论是计算机科学最重要的组成部分。根据计算复杂性理论，自然界和科学理论上的所有问题，可以分为三大类[1]：多项式问题，即可以在多项式时间内得到解答的问题；指数型问题，即可以在指数型时间内得到解答的问题；不可解问题，无论花多少时间，都不可能得到问题的解答。对前两类问题，我们可以很容易找到确切的例子来进行说明，但对于最后一类，我们则很难找到具体的例子来说明这个例子就是不可解问题。这里请注意，不可解问题与问题本身无解是两个不同的概念。例如，我们可以很容易地设计出一个方程，使得该方程无解，我们也能很容易地判断该方程无解。而不可解问题则指的是，问题本身可能是有解的，只是我们可能无论花多长时间都得不到这个解。迄今已发现了一些问题是不可解的，但是，这些问题是否真的不可解则大有疑问。例如，停机问题就被证明是不可解的，但它的证明存在着很大的争议。著名的 3x+1 问题到目前为止亦被认为是不可解的，或至少到目前是解不出来的[2]。同样，谁也不能肯定它就永远不可解。

关于不可解问题，历史上有两大权威，代表着两个绝然对立的观点。一个是二十世纪最伟大的数学家希尔伯特，另一个就是以发明了不完备定理而闻名于世的哥德尔。哥德尔的不完备性定理指出，在某个数学公理系统中，总有不可证明的命题存在。公理系统的这种性质叫做不完备性。从哲学上，哥德尔定理在某种程度上体现了不可知论。而希尔伯特的著名命题则是：我们必须知道，我们必将知道。也就是说，他认为任何数学问题最终必能得到解决。必须指出的是，尽管哥德尔的不完备定理被称为“定理”，但由于它的证明沿用了康托尔的对角线法[3]，而康托尔的对角线法一直是存在着很大争议的。所以，哥德尔的那个“定理”是否真的就有效尚有疑问。本文或笔者所持的观点是：倾向于赞同希尔伯特。

3x+1 问题与许多科学领域，特别是一些数学领域，具有直接的关联，如：数值理论，计算理论，概率论，离散动态系统[4]，计算机科学[5][6]，等。

因此，3x+1 问题研究具有重要意义，一是它的任何突破直接对上述领域的研究发展具有意义，二是它能直接将不可解问题的研究引向深入。

## 2 3x+1问题概述

输入为任意一个自然数，若该数为奇数，将它乘以3然后加1，结果继续；若是偶数，就将其除以2，结果继续，直到得到1为止。现在的问题是：是否对所有的自然数，结果都能得到1。这就是著名的3x+1问题，迄今没有确定的解答。

从一个自然数开始，用上面这个变换，我们可以计算出一串自然数的序列。为了形象起见，通常把这串数列叫做以最初用来开始计算的那个自然数命名的“航班”。比如说，第6次航班就是6→3→10→5→16→8→4→2→1。我们把一个航班里的最大数字，叫做这个航班的“最大飞行高度”。比如说，第6次航班的最大飞行高度就是16。我们把航班在数字1“着陆”之前的数字个数（最初的数字包含在内，但1不包含在内），叫做这个航班的“航程”（特别定义第1次航班的航程为0）。第6次航班的航程就是8。如果真有自然数在此变换下永远达不到1，那么这个航班

的航程就是无穷大了。注意，航程又称“完全停止时间”(total stopping time)[7]。

我们把从起点开始直到第一次遇到小于起点的数为止，所需的变换次数，叫作“停止时间”(stopping time)[7]。不妨举一个例子来说明：第11次航班是11→34→17→52→26→13→40→20→10→5→16→8→4→2→1。我们看到从起点开始，34，17，52，26，13，40，20都不小于起点11，第一次遇到的小于11的数是10，到达10需要8步，所以第11次航班的停止时间为8。一个最简单的推论就是，偶数次航班的停止时间总是1，因为一开始就除以2，只需要1步就跌到较低的高度去了。为什么我们对一个航班的停止时间感兴趣呢？因为如果所有航班的停止时间都是有限的话，3x+1问题就成立了。让我们假设已知所有航班的停止时间都是有限的，用数学归纳法来证明3x+1问题，也就是所有的航班都在1上“着陆”。我们已经知道第1到第8航班都是在1上着陆的，现在假设对于所有小于n的数字k，第k次航班都在1上着陆，我们来看看第n次航班的情况：由于按假设它的停止时间是有限的，所以它迟早会降落在一个比n小的数字上——于是按归纳假设它就会降落在1上！

下面要说说几个记录，目前3x+1问题已经被检验到$20*2^{58}$，都没有发现反例。这是葡萄牙阿弗罗(Aveiro)大学的Tomas Oliveira e Silva的工作[12]，他用了很巧妙的编程方法。如果一个航班的航程大于所有它前面的航班的航程，我们就把它叫作“航程纪录航班”，比方说第7航班，它的航程是16，比第1到6次航班的航程都长，所以第7航班是个航程纪录航班。今天我们已经知道的航程纪录航班有118个，航程最长的是2234047405400065次航班，它的航程是1871，这是Eric Roosendaal发现的，他有个个人网站http://personal.computrain.nl/eric/wondrous/，里面有各种各样关于3x+1问题的信息，下面的记录也都来自这个网站。

同样的，如果一个航班的停止时间大于所有它前面的航班的停止时间，我们就把它叫作“停止时间纪录航班”，比方说从上面的表中我们看到第7航班也是个停止时间纪录航班。今天已知的停止时间纪录航班有30个，航程最长是1008932249296231次航班，它的停止时间是1445。最大飞行高度记录航班就是那些最大飞行高度记录大于所有它前面的航班的那些航班，现在已知的有76个，最大的是10709980568908647次航班，到达了350589187937078188831873920282244的高度。对于一个固定航班N，考虑它在1着陆之前所作的变换，如果把其中除以2的变换称为“偶变换”并记为E(N)，而把乘以3再加1的变换称为“奇变换”并记为O(N)。数学家已经证明，O(N)/E(N)<log2/log3。

我们注意到，对有些航班来说，O(N)/E(N)非常接近于log2/log3≈0.63092975……。有猜想认为它会越来越接近这个数字（也有相反的猜想，认为不会无限接近），所以大家为此设立了另一个纪录，就是这个比值比所有以前的航班更接近log2/log3的航班。这样的纪录不多，现在已知的有15个，其中最后一个是N=100759293214567，I(N)/P(N)≈0.604938。值得一提的是N=104899295810901231，它的这个比值还要更靠近，达到0.605413，但是我们不知道它是否是一个纪录，也就是说，我们不知道所有比它小的航班里，是否还有比这个比值更靠近log2/log3的。

我们知道，对于任何p，总有至少一个航班，它的航程是p：$2^p→2^{p-1}→2^{p-2}→……→4→2→1$ 但是一般并不需要这么大的航班，就可以达到航程p。在2000年有人提出要找到最小的航班号，使得它的航程恰好是2000。现在最好的纪录是第67457283406188652次航班，但谁都不知道这是不是最小的航程为2000的航班。

计算一个航班的算法是非常简单的——只要除2或乘3加1。但是为了检验大量的和航次巨大的航班，巧妙的编程方法是非常重要的。上面的那些纪录都是由几台类似于我们平时使用的那样的计算机得到的结果。但是如果没有好好地思考和编程，光是硬算，那就很难得到这样的结果。为了验证一个航班的确在1上着陆，并不一定需要把结果计算到1。如果你已经验证了所有航次小于n的航班都在1上着陆，那么对于第n次航班，你只要把结果计算到一个小于n的数m就可以了——我们已经验证过第m次航班在1上着陆。事实上，如果我们只要计算到一个以前的航班飞行时到达过的数值就可以了，当然这需要记住以前已经到达过的比较高的高度，这里也必须巧妙地编程使得这样的记忆所使用的内存比较少。更重要的是使用数学方法去减少计算量。比如说，任何n=4k+1的航班最终都会飞到一个比n更小的高度。首先这是奇数，我们乘3加1得到

12k+4，然后连除两次2，就有3k+1<n。所以我们没有必要费功夫去验证4k+1型的航班。另外偶数次航班第一次变换就被除以2，降低了高度，所以同样也不需专门验证。只用这样一个小技巧，我们就使计算量减少到原来的25%。如果按照这样的思路下去，我们同样不需要考虑16k+3型的航班，只要考虑到前面的飞行记录：
16k+3→48k+10→24k+5→72k+16→36k+8→18k+4→9k+2→……而9k+2<16k+3。我们可以这样追踪下去，考虑256k+i型的航班，其中i取0到255，那么我们会发现我们需要考虑的类型只有i=27、31、47、63、71、91、103、111、127、155、159、167、191、207、223、231、239、251、255。这样我们要作的计算只有最初的8%不到。而Eric Roosendaal得到上面那些纪录的程序，是建立在对65536k+i型航班分析的基础上的，其中只有1729种航班需要真正的检验（只有原来计算量的2.6%）。他的程序还使用了其它的算术技巧，以及可以同时计算好几个航班。Tomas Oliveira e Silva进一步改进了这些技巧，从而使得他成为现在3x+1问题验证的世界纪录保持者（他的计算从1996年8月开始，到2000年4月结束，其间使用了两台133MHz和两台266MHz的DEC Alpha计算机）。Eric Roosendaal还在和其他人一起合作进行计算（包括再次验证以前的结果），如果谁愿意加入这个研究项目的话，可以去访问上面给出的他的主页。

## 3 国际上已经取得的一些进展

R. Terra[7][8]和C. Everett[9]证明了，“几乎所有的航班都会下降到它的起始点以下”，也就是说“几乎所有航班的停止时间都有限”。这里的“几乎所有”是有确定的数学意义的，它是指：
——存在一个自然数$n_1$，在所有小于$n_1$的航班里，最多只可能有1/10的航班，它们的停止时间无限；
——存在一个自然数$n_2$，它比上面的$n_1$要大，在所有小于$n_2$的航班里，多只可能有1/100的航班，它们的停止时间无限；
——存在一个自然数$n_3$，它比上面的$n_2$要大，在所有小于$n_3$的航班里，最多只可能有1/1000的航班，它们的停止时间无限；——等等等等……
这好象很接近证明“所有的航班的停止时间都有限”了，于是很接近证明猜想本身了。但是好好想想，这个结论只不过是说明停止时间无限的航班会越来越稀少罢了，它们还是有可能存在的，并且，这个结论一点也没有排除有其它循环存在的可能。

对于在1上着陆的航班，数学家们也得到了一些结果。他们证明了，存在一个常数c，当n足够大的时候，在比n小的航班中，能够在1上着陆的航班的个数大于等于$n^c$。在1978年R. Crandal首先给出c=0.05，虽然小了点，但毕竟是开头一步；然后J. Sander给出c=0.3；在1989年I. Krasikov得到c=0.43；1993年G. Wirsching得到c=0.48；1995年D. Applegate和J. Lagarias得到c=0.81。目前最好的结果是I. Krasikov and J. Lagarias于2002年得到的0.84[10]。看起来我们越来越接近c=1这个最终目标了。可是我们不知道现在用来得到c的方法是否还可以再用下去以最终达到1。

1995年的这个证明相当特殊。它使用了计算机程序来解一个十分巨大的方程组，所以这个证明不能用手工来验证。在论文中，我们看见的不是一个关于c=0.81的定理的证明，而是一个关于如何写出这个巨大方程组的说明，和由程序计算出来的结果，以及如何使用这些结果来解释c=0.81。其他的数学家如果想验证这个结果，必须首先看懂关于方程组的证明和那些解释，再按照里面的说明来写一个程序，运行它，再看看结果是否和文章中的相同。还有一些结果是关于如果有其他不同于4→2→1的循环存在时，对这样的循环的性质的研究。R. Crandal和N. Yoneda在1978年证明，如果这样一个另外的循环存在的话，那么它的长度（就是在这个循环中数字的个数，比如说循环4→2→1的长度就是3）一定要大于275000。1993年这个体积增大到17087915，最近的结果是102225496。这些结果是通过分析包括我们前面提到的各种纪录得到的，所以这些结果我们还是不能完全通过手工来验证。我们看到，如果真有另外的循环存在的话，那一定是非常非常巨大的！

数学中有一种叫“启发式”的论证方法，建立在估计和概率的手段上。比如说底下的论证方法就是这个类型的：

“每个数字要么是奇数要么是偶数，如果随便取一个自然数，碰到奇数和偶数的可能性是一样的。如果我们把一次航班中这一系列数值看是随机的话，那么使用奇变换和偶变换的可能性也是一样的，所以平均在每两次变换中我们有一次是n→3n+1，有一次是n→n/2。所以平均起来，每次飞行高度的变化就是乘以3/2，于是……就会越飞越高。”

这样的启发式论证就推翻了原来的猜想!但是这个论证显然是错误的，因为它没有考虑到，每一次奇变换后随即而来的一定是一次偶变换，因为如果n是奇数的话，3n+1一定是偶数；而每一次偶变换后随即而来的却不一定是一次奇变换。J. Lagarias改进了这个启发式论证。他指出，如果我们把奇变换后再作偶变换考虑在一起，那么这样得到的结果可以看作是真的“很随机”。于是有1/2的可能性它是奇数，有1/4的可能性是一个奇数的2倍，有1/8的可能性是一个奇数的4倍，等等。于是飞行高度的变化就是以下变换的“平均效应”；

——n乘以3/2，这有1/2的可能（奇变换后再作偶变换的结果为奇数）；

——n乘以3/4，这有1/4的可能（奇变换后再作两次偶变换）；

——n乘以3/8，这有1/8的可能（奇变换后再作三次偶变换）；

…………

于是平均来讲，每次变换后高度的变化就是$c=(3/2)^{1/2}(3/4)^{1/4}(3/8)^{1/8}(3/16)^{1/16}\cdots\cdots=3/4$

所以高度在总体上来说应该是越来越低，每次大约低25%，最终降到一个循环上（不过这个论证没有排除有除了4→2→1以外的其他循环）。这个论证可以使我们使用论证中的模型来计算出，从一个自然数开始，平均要多少步的这样的飞行（就是停止时间中奇变换的次数），可以使飞行高度降到起始点以下。S. Wagon推断出[11]，随着自然数的增大，停止时间的平均值理论上将趋向于一个常数，约为：9.477955。他并对3到$10^9$之间的航班的停止时间的平均值进行了验证，结果与上述数值接近。这两个结果惊人的一致性使我们相信上面的启发性模型是正确的。如果它是正确的，那么就意味着没有停止时间无限的航班，于是3x+1猜想就是正确的，至少可以得出没有飞得越来越高的航班的结论。可是一个启发性论证，就算再有实验证据来表明它是对的，也只不过是个论证，只能使我们对猜想的正确性更充满信心。它不能代替真正的数学证明。用这个启发式的概率模型，我们还可以预言，对于第n次航班，它的最大飞行高度不会超过$Kn^2$（对于某个常数K）。数值计算表明对于K=8，这个公式是正确的（同样地，这可以让我们提出猜想，而不是证明定理）。

## 4　本文的解决方案

先定义几个概念，如前所述，对于任一自然数，经过多次变换后，第一次得到比该自然数小的结果，所需的计算步骤，称为停止时间m(stopping time)[7]，其中乘以3然后加1的次数，称为停止时间奇变换次数$m_1$，除以2的次数，称为停止时间偶变换次数$m_2$，显然，$m=m_1+m_2$。注意，我们的这个停止时间，指的是从最初那个数开始，直到第一次遇到比该数小的自然数，所需要的计算步骤，即乘以3加1和除以2的计算次数。

我们假定3x+1问题的猜想成立，即所有自然数经连续变换后，最终都会回到1。在此前提下，有两种可能性。一种可能是，当自然数足够大时，其停止时间m以及$m_1$、$m_2$均达到一个极限值，以后不会再增加。根据前一节的启发式证明，这种假设极可能成立。第二种可能性就是，对每一自然数，尽管最终都能降到比该自然数小，但其停止时间m以及$m_1$、$m_2$会随着自然数的增大而不断增大。由于前一节仅仅只是启发式思考，并非严格的证明，故这种可能性也不能完全排除。本文尊重上一节的启发式思考，主要针对在第一种可能性下的解决方案。

定理1：若存在某个自然数N，对于所有小于$2^N$的自然数，它们的停止时间偶变换均不超过N，则对于任意大于$2^N$的自然数，其停止时间偶变换也不会超过N。也就是说，若我们能找到这样的N，则意味着证明了3x+1猜想。

证明：偶自然数就不用证明了，所有大于$2^N$的奇自然数，均可表达为：$i+k_0 2^N+k_1 2^{N+1}+\cdots\cdots$，

其中i为小于$2^N$的奇自然数，$k_0$, $k_1$……要么为0，要么为1。对该数进行变换，注意我们仅对i进行3x+1或除以2的变换，而对后面的数要么乘以3要么除以2，与前面同步进行，当i这一边达到了小于i时,显然此时后面的数也会达到比它的最初值小，从而整个数都比它的最初值小。由于i的停止时间偶变换不超过N，故整个数的停止时间偶变换亦不超过N。而当停止时间偶变换不超过N时，停止时间也会不超过一个常数。从而，所有自然数的停止时间不超过一个常数，按归纳法，3x+1猜想成立。证毕。

显而易见，若能找到这样的N，3x+1问题也就彻底解决了。但，如前所述，目前已发现的最大停止时间达1445，所对应的停止时间偶变换接近900，而要验证$2^{900}$内的所有自然数，人类现有计算机的计算能力远远不够。

进一步研究，我们发现，有一种方法可以大大缩小验证的计算量。对于某个自然数k，我们假定对于所有小于$2^k$的自然数，绝大部分其停止时间偶变换都小于等于k，只有极少数不是这样。我们假设不符合小于等于k这一规则的数有x个，那么，对于$2^k$至$2^{k+1}$之间的数，也只有这x个数再加$2^k$有可能不符合规则，其他的都符合。进一步，对于$2^{k+1}$到$2^{k+2}$之间的所有自然数，那就会有x个数加$2^{k+1}$，以及x个数加$2^k$再加$2^{k+1}$，也就是有2x个数有可能不符合上述规则，依此类推，随着k的递增，有可能不符合规则的数将按2的倍数递增。但是，另一方面，我们注意到，当k增加1时，x个数里面有一部分会从不符合规则变为符合规则，也就是说，对于某个数i，它的停止时间偶变换大于k，但$i+2^k$的停止时间偶变换很可能不超过k+1，从而由不符合上述规则，变为符合规则。

综上所述，随着k的增大，不符合上述规则的个数有两个趋势，一个趋势增大，另一个趋势减小。我们只要找到这样的规律，自某个数k开始，减小的趋势快于增大的趋势，就可证明，最终所有数都符合上述规则。上述增大的趋势是以2的倍数增大，也就是说，只要自某个数k开始，对于小于$2^k$的数，至少有一半不符合规则的数，当加上$2^k$之后，其停止时间偶变换不超过k+1，只要发现了这样的变化规律，就意味着彻底解决了3x+1问题。

注意，考虑到当前计算机的计算能力，找到“自k开始，对于小于$2^k$的数，至少有一半不符合规则的数，当加上$2^k$之后，其停止时间偶变换不超过k+1”这样的k可能较难，也就是说这样的k可能较大，导致计算量超大。但这个问题可以在很大程度上得到解决，我们只要能发现上述减小的趋势，其减小的幅度在不断上升，最后能达到或超过二分之一，这样一个规律，也就能证明我们所要的结果。

根据前一节所述，停止时间的平均值趋向于一个常数[11]，同样，停止时间偶变换的次数也具有这一性质。这就意味着，随着k的增大，不符合上述规则的数的个数，其数学期望将趋向于零，也就是说，这样的数的个数，或者为零，或者小于某个常数。因此，若按照我们的方法，通过使用大型计算机以及互联网的协作计算能力，3x+1问题极可能很快得到彻底解决。

## 5 结束语

3x+1问题牵涉到许多研究领域，同时由于该问题被许多权威专家判定为超难，它的成功解决不仅会对许多相关领域的研究产生推动促进，更可能意味着方法论上的重大突破。本文通过深入分析3x+1问题随着其输入自然数的不断扩大所蕴含的演化规律，给出了一条有力可行的解决该问题的方案，随着大型计算机以及互联网计算功能的应用，本方案可能直接导致决定性结果。

### 参考文献